\documentclass[aps,preprintnumbers,amsmath,amssymb,prd,superscriptaddress,longbibliography,twocolumn,
floatfix,nofootinbib]{revtex4-2}
\usepackage{graphicx}
\usepackage{dcolumn}
\usepackage{color}
\usepackage[caption=false]{subfig}
\usepackage{feynmp}
\usepackage{feynmp-auto}
\usepackage{float}
\usepackage{systeme,mathtools}
\usepackage{stmaryrd}
\usepackage{easyReview}

\def\comment#1{}

\usepackage{bm,latexsym,mathrsfs,enumerate,color}
\usepackage[mathcal]{euscript}
\usepackage[breaklinks=true,unicode=true,urlcolor = blue,colorlinks = true,citecolor = blue,linkcolor = blue]{hyperref}
\usepackage{graphicx}
\usepackage{todonotes}
\usepackage{wrapfig}

\renewcommand{\vec}[1]{\bm{#1}}

\def\slashchar#1{\setbox0=\hbox{$#1$}           
	\dimen0=\wd0                                 
	\setbox1=\hbox{/} \dimen1=\wd1               
	\ifdim\dimen0>\dimen1                        
	\rlap{\hbox to \dimen0{\hfil/\hfil}}      
	#1                                        
	\else                                        
	\rlap{\hbox to \dimen1{\hfil$#1$\hfil}}   
	/                                         
	\fi}                                         %

\DeclareMathAlphabet\mathbfcal{OMS}{cmsy}{b}{n}

\def\nablab{{\mbox{\boldmath $\nabla$}}}

\def\gammab{{\mbox{\boldmath $\gamma$}}}

\def\Omegab{{\mbox{\boldmath $\Omega$}}}

\begin{document}
	
\title{Bose-Einstein condensation and superfluidity on a fuzzy sphere}

\author{Vira Shyta}
\affiliation{Max Planck Institute for the Physics of Complex Systems, N\" othnitzer Straße 38, 01187 Dresden, Germany}

\author{Flavio S. Nogueira}
\affiliation{Institute for Theoretical Solid State Physics, IFW Dresden, Helmholtzstr. 20, 01069 Dresden, Germany}

\author{Ashley M. Cook}
\affiliation{Max Planck Institute for the Physics of Complex Systems, N\" othnitzer Straße 38, 01187 Dresden, Germany}
\affiliation{Max Planck Institute for Chemical Physics of Solids, N\" othnitzer Straße 40, 01187 Dresden, Germany}

\date{Received \today}

\begin{abstract}
According to Hohenberg's theorem, Bose-Einstein condensation (BEC) in two dimensions is impossible for any temperature $T>0$. 
By contrast, superfluidity does occur in two dimensions at finite temperatures; it emerges due to the breaking of Galilei invariance. Here we consider both BEC and superfluidity on a compact two-dimensional space taking the form of a non-commutative (``fuzzy") sphere, where the scalar bosonic fields are promoted to $N\times N$ matrices. Here, $N$ is directly related to the non-commutativity parameter of space and introduces an additional scale into the system. We find that non-commutativity favors ordered phases and so enhances BEC and superfluidity.
We analyze BEC in ideal and weakly interacting Bose gases on a fuzzy sphere, finding in each case that the critical temperature of BEC is greater compared to that found in the case of an ordinary (commutative) sphere $S^2$. Then we investigate the superfluid response of weakly interacting Bose systems. 
To account for vortices that can occur in a superfluid, we show that, even on an ordinary sphere, the collective coordinates of vortices induce non-commutativity. With this in mind, we proceed to  expand the definition of vortex defects to an inherently non-commutative sphere studied here, where the notion of a point is untenable.  
The non-commutativity is expected to be experimentally relevant to both BEC and superfluidity since the fuzzy sphere possesses  a  thermodynamic limit distinct from the one defined over a plane, unlike the  $S^2$ case.
 The significance of this difference is illustrated by the superfluid density calculation indicating that, in the large sphere limit, the normal fluid fraction  on the fuzzy sphere yields a linear in $T$ dependence, while on a commutative  $S^2$ it exhibits the usual two-dimensional $\sim T^3$ behavior. This linear dependence, which is a direct consequence of non-commutativity, is reminiscent of Uemura's law in cuprate high-$T_c$ superconductors.    	
\end{abstract}

\maketitle

\section{Introduction}
\subsection{Bose-Einstein condensation  and superfluidity in two dimensions}

The phase structure of a $U(1)$-invariant, homogeneous non-relativistic Bose system consists typically of a superfluid and a normal fluid phase. Physical properties of a superfluid  cannot be explained by classical physics, and for this reason it is dubbed a quantum liquid \cite{annett2004superconductivity}. 
 While a phase structure of a Bose system can, under certain conditions, involve, for instance, a Mott insulating phase~\cite{Fisher_PhysRevB.40.546} or a supersolid phase~\cite{Balibar_supersolid}, already the superfluid to normal-fluid phase transition is characterized by nontrivial behavior. This is most evident in two dimensions:
 the Mermin-Wagner theorem \cite{Mermin-Wagner_PhysRevLett.17.1133}  forbids spontaneous symmetry breaking in such a system, yet the phase transition is still possible via the celebrated  Berezinskii-Kosterlitz-Thouless (BKT) phase transition~\cite{berezinskii1971destruction,kosterlitz1973ordering} driven by a vortex-pair unbinding mechanism. 
The long-standing interest in two-dimensional Bose systems is sustained by the fact they provide realizations of several unconventional quantum states of matter, such as high-temperature superconductivity \cite{CHEN20051, Carbotte_2011, DAS20261354908, BKT-cuprates-2020, BEC-BCS-SC-review-2024} and quantum critical planar magnets \cite{,BKT-frust-mag, ZHANG2025100256}. There are also engineered quantum systems studying the BKT transition experimentally; prominent examples include ultracold atoms  \cite{BKT-ultracold-2006, PhysRevLett.128.250402, BKT-ultracold-2025} and Josephson junction arrays \cite{2DJJAs,tinkham2004, Marcus-2018, PhysRevB.102.094509,Marcus-2024}.

The BKT transition occurs, however, in absence of Bose-Einstein condensation (BEC), which is forbidden for $d=2$ at finite temperature, $T>0$, as dictated by Hohenberg's theorem \cite{Hohenberg_PhysRev.158.383}, a close relative of  the Mermin-Wagner theorem. This highlights the fact that BEC is not necessary for superfluidity to occur. On the other hand, a homogeneous ideal Bose gas in $d>2$ exhibits BEC for finite temperatures below a certain critical temperature, $T_c$, yet it is not, strictly speaking, a superfluid (although sometimes we speak of an ``ideal superfluid'')~\cite{pitaevskii2003bose}. The reason behind this is the Landau criterion for superfluidity, which posits that frictionless flow of a quantum liquid requires the breaking of Galilei invariance. Meanwhile, the ideal Bose gas is Galilei invariant in the BEC phase. 
 
 There are two-dimensional Bose systems in which BEC becomes possible at $T>0$. Instead of focusing on an infinite two-dimensional plane, one can study BEC and superfluidity on surfaces of finite area. One possibility is to consider an inhomogeneous space, as in the case of two-dimensional ultracold atomic Bose gases in either a magnetic or an optical trap, which is necessarily finite  \cite{pitaevskii2003bose}. For instance, in the case of a harmonic trap, the obstruction to BEC occurs at $d=1$ rather than $d=2$. Alternatively, preserving the homogeneity of the two-dimensional space, one could instead consider a Bose gas on a sphere.   In such configuration,  the Bose gas also undergoes BEC at finite temperature~\cite{Tononi2019}. There is considerable interest in experiments involving ultracold gases on general two-dimensional curved geometries \cite{shell-shaped-BEC-2018, BEC-curved-2020,Tononi-Review}. One motivation to study this type of system is the experimental prospect of trapping ultracold atoms in a geometry that is not flat, like, for example, a bubble type of trap, that can be achieved in microgravity conditions  \cite{Tononi_PhysRevResearch.4.013122}. Some experiments where the effect of gravity has been compensated were realized in the past \cite{Tononi-Review}.

In this work, we investigate BEC and superfluidity  in a two-dimensional non-relativistic Bose system that, crucially, resides on a non-commutative (also known as ``fuzzy'') two-dimensional sphere.  One of the remarkable differences between such a system and its commutative-sphere counterpart  lies in the nature of the thermodynamic limit.
In the commutative case, taking the large-radius limit required by thermodynamic limit may lead to results that have locally the same behavior as a system defined on $\mathbb{R}^2$.   
We will show that systems on a fuzzy sphere   retain the effects of curvature in the thermodynamic limit. 

For the sake of clarity in presenting the outline and main results of this paper, it is useful to first introduce the essential concepts of non-commutativity as an inherent  quantum property of  matter. This also provides a physically-motivated framework for the fuzzy sphere construction.

\subsection{Non-commutativity in condensed matter}

As already mentioned, here we set out to study  properties of a Bose gas on a \textit{fuzzy} sphere~\cite{Madore_1992}.  
From quantum mechanics, we know of an intrinsic connection between two non-commuting operators and an ensuing uncertainty relation. In the  case of a fuzzy sphere, it is the spatial coordinates that do not commute. Consequently, in view of the arising uncertainty in knowledge of the position coordinates, the concept of a point in space is not strictly applicable anymore and, hence, one can think of the sphere as being fuzzy. This can be understood as an emergent discreteness of space, where non-commutativity introduces an additional length scale, in a way analogous to the role the Planck constant plays in  Heisenberg's uncertainty principle.  
It is exactly the influence of this additional length scale---a property of the space to which the Bose gas is confined---on the BEC and superfluid behavior that we want to explore.

There are quite a few reasons to explicitly consider non-commutativity of space  in this condensed matter problem. To begin,  non-commutativity is inherent when considering particles moving on a two-sphere~$S^2$.  This follows from an interesting property of any such particle: the only momentum available to it is angular momentum. 
Take, for example, the Hamiltonian of a free particle, $H_{\rm free}=(m/2)\vec{v}^2$. Since $\vec{r}^2=R^2$ (note that although the notation $S^2$ is normally used to denote a unit sphere, we are assuming here that the radius $R$ is any arbitrary number), we obtain that $H_{\rm free}$ can be rewritten as $H_{\rm free}=\vec{L}^2/(2m R^2)$, where $\vec{L}=\vec{r}\times m\vec{v}$ is the angular momentum. Quantum mechanically, we are led to consider a canonical momentum operator of the form, $\vec{p}=\vec{L}/R$.  Since angular momentum components satisfy an SU(2) Lie algebra, the equivalence between momentum and angular momentum implies that different momentum components, $p_i$ and $p_j$, do not commute. More specifically, one obtains the non-commutativity relation, $[p_i,p_j]=(i/R)\epsilon_{ijk}p_k$,  where $\epsilon_{ijk}$ is the Levi-Civita tensor.  
In this case, $1/R$ can be viewed as a non-commutativity parameter, highlighting the fact that this is a spherical-geometry effect not occurring on an $\mathbb{R}^2$  plane. 

While these arguments illustrate how the fuzzy sphere is natural in consideration of quantum systems at a single-particle level, it is also well-motivated by many-body physics.
The first experimental confirmation of the BKT transition was made possible by absorbing a thin superfluid helium film on an oscillating substrate \cite{Reppy_PhysRevLett.40.1727}. A spherical substrate could in this case provide a platform to probe superfluidity on a sphere. The oscillating substrate introduces a characteristic frequency that provides a natural cutoff to the angular momentum modes associated to the energy excitation spectrum of the superfluid. As we will discuss later, such a cutoff arises naturally as an additional length scale on a fuzzy sphere, where it is associated to the non-commutativity parameter of space~\cite{balachandran2006, Nair_PhysRevD.85.045021,Polychronakos_PhysRevD.88.065010}. 
Moreover, when observing a vibrating spherical substrate, coordinate points on it may appear ``blurred", which would also influence the appearance of vortices in this superfluid, making them different from the point-like objects widely used as approximations to describe vortices on the $\mathbb{R}^2$ plane. This behavior can be viewed as a reflection of some uncertainty relation between the coordinates $x_i$ of $\vec r \in\mathbb{R}^3$ satisfying the constraint $\vec r^2=R^2$.
The fuzzy sphere offers a way to  capture the non-commutative character of the underlying geometry of this setup.

Non-commuting coordinates like the ones alluded to in the preceding paragraphs naturally occur in the lowest Landau level (LLL) projection associated to the quantum Hall effect \cite{Pasquier2007,Haldane_PhysRevLett.51.605,Fano_PhysRevB.34.2670,Greiter_PhysRevB.83.115129, susskind2001_qh, polychronakos2001quantum,SimeonHellerman_2001, haldane2016geometrylandauorbitsabsence, goldman2022}. Related non-commutativity emerges in nonlinear sigma models via central extension of the momentum commutator~\cite{HitoshiCentralExt} particularly relevant to the quantum Hall ferromagnet~\cite{sondhi1993}. This also motivates the study of fuzzy nonlinear sigma models~\cite{balachandran2004}. Furthermore, the fuzzy sphere has been quite important as a regularization strategy to investigate critical phenomena in 2+1 dimensions.  Similarly to the way we will introduce non-commutativity on a sphere via the LLL projection, 
	some particularly useful numerical methods \cite{zhou2025fuzzifiedjuliapackage,Fuzzy-CFT-annurev} in conformal field theories  employ the fuzzy sphere formalism with a monopole as a source of the magnetic flux and construct bosonic fields via composite fermions \cite{Fuzzy-CFT-annurev,He_PhysRevX.13.021009}. Hence, the fuzzy sphere has more generally been explored as an efficient computational  tool, for example, for characterization of the 3D Ising  model~\cite{hu2023} and nonlinear sigma models~\cite{araz2023}.

Moreover, non-commutativity related to the LLL physics  arises in rotating superfluids \cite{Son-Tkachenko-mode-noNC2018, Son-NC-2024, Manoj2025, glodkowski2026}.  One way to see this is to consider the emergent quantum vortex lattice excitations, known as Tkachenko waves \cite{tkachenko1966vortex, tkachenko1966stability, tkachenko1969elasticity}, in a superfluid under rotation. The fact that an effective field theory of Tkachenko modes is non-commutative and can in some instances be viewed as the LLL projection is not a surprise due to the duality between a neutral particle under rotation and a charged particle in the magnetic field \cite{BECinLLL2021,Senthil-rotatingSF-LLL-2024}. However, this highlights the relevance of the fuzzy sphere treatment to rotating superfluids and supersolids \cite{Moroz2020-SciPost}, especially in the context of experimentally-realized rotating Bose-Einstein condensates in the LLL \cite{RotatingBEC-LLL-2004, BECinLLL2021, BECinLLL2022}.

\subsection{The fuzzy sphere construction}

We turn now to the construction of the non-commutative space using an approach found in many works focused on the implications of fuzzy sphere geometry~\cite{Aschieri:2003vy, Aschieri:2004vh, Madore_1992,Balachandran-2002,Nair_PhysRevD.85.045021,Polychronakos_PhysRevD.88.065010,gavriil2015higher, balachandran2006, hasebe2024, Hasebe:2010vp, Hasebe:2014nia}. Let us consider a magnetic monopole located in the center of a ball of radius $R$ whose boundary is $S^2$. The monopole produces a magnetic field perpendicular to every point on the sphere. The angular momentum of a charged particle moving on $S^2$ under the influence of a monopole flux is given by,
\begin{equation}
	\label{Eq:L-monopole}
	\vec L=\vec{r}\times m\vec v+eg \frac{\vec r}{R},
\end{equation}                         
where $e$ and $g$ are the electric and magnetic charges, respectively, and we have set the speed of light $c=1$. Thus, a point at rest corresponds to $\vec L=eg\vec r/R$. Duality and flux quantization implies that $eg=N/2$, $N\in\mathbb{N}$, so quantum mechanically we replace the spatial coordinates $x_i$ by operators $X_i$ to naturally obtain the Lie algebra of non-commuting coordinates, 
\begin{equation}
	\label{Eq:Space-Commut}
	[X_i,X_j]=i\frac{2R}{N}\epsilon_{ijk}X_k,
\end{equation}
where we have defined, $X_i=(2R/N)L_i$. Each $L_i$ corresponds to the $N$-dimensional representation of the angular momentum operator, so each $L_i$ is represented by a $N\times N$ matrix. The angular momentum quantum number $l$ assumes here a maximum value $M=N-1$. Throughout this work, our fuzzy sphere will be defined by  the commutation relation in Eq.~\eqref{Eq:Space-Commut}.

Since we aim to study Bose systems on the fuzzy sphere, we need to define a scalar field theory over such space. Adopting the approach discussed in Refs.~\cite{Nair_PhysRevD.85.045021,Polychronakos_PhysRevD.88.065010}, we promote the bosonic scalar fields to Hermitian matrices of dimension $N\times N$. In this case, derivatives are defined via a Lie derivative of the form, $\mathcal{L}_iF=-i[L_i,F]$, where $F$ is a matrix field. This can be motivated by a  simple argument from matrix quantum mechanics. Namely, the time evolution of a matrix $M$ is given by $dM/dt=-i[H,M]$, where the Hamiltonian $H$ is viewed as the generator of time translations. Thus, we could think of a time derivative as being simply given by $-i[H,M]$. Similarly, we use the generators of rotations to define the ``partial derivatives", $\partial_i F\to -i[L_i,F]$ \cite{MADORE1992187}, and this allows us define the Lie derivative, $\mathcal{L}_iF=-i[L_i,F]$. The next  ingredient necessary to adapt the scalar field theory to the fuzzy sphere formalism is the counterpart to  integration over the sphere. Since one works with matrix fields and the derivatives are given by commutators, the integral $\int_{S^2}d^2x$ is replaced by the matrix trace  $(4\pi R^2)/N\operatorname{Tr}$  with an additional prefactor needed to preserve the dimensionality. 

We note that variation of $N$ has been studied as part of procedures for dimensional reduction~\cite{Zhang:2001xs, Bernevig:2002eq}, motivated in part by the essential role of non-commutativity in Banks-Fischler-Shenker-Susskind matrix theory~\cite{Banks:1996vh}. Certain simplifications in treatment of non-commutativity are possible for large $N$, which might naively be associated only with a commutative limit, employing the star product~\cite{dong2020, goldman2022}.  Interestingly, matrix dimension $N$ has also been associated instead with the number of spinless electrons occupying the LLL as part of a matrix theory treatment of Laughlin quantum Hall states~\cite{susskind2001_qh, polychronakos2001quantum,SimeonHellerman_2001}, though this generally introduces further deformations of the fuzzy sphere~\cite{patil2025microscopic}. This more complex framework has included study of vortices, where the  non-commutativity was associated with generalization beyond the point-limit ~\cite{tong2015,dorey2016,doreymatrix2016}. This treatment is also of interest in light of the recent identification of an unconventional pairing mechanism~\cite{patil2025microscopic, banerjee2024} potentially relevant to superconductivity.

\subsection{Outline of the paper}

Having put Bose systems into the non-commutativity context and introduced the basic building blocks for our field theory on a fuzzy sphere, we present the plan of the paper. Section \ref{Sect:BEC-fuzzy} is dedicated to BEC in an ideal Bose gas on a fuzzy sphere. 
There we show that fuzziness increases the critical temperature $T_\text{BEC}$ below which the ideal gas is condensed. 
Crucially, this section also provides a proper definition of the thermodynamic limit on a fuzzy sphere, since it is required to precisely define BEC. 
Section \ref{sec:intBEC} considers a fuzzy weakly interacting  Bose system within the Bogoliubov approximation. The standard calculation to be done in this case is the so-called depletion of the condensate \cite{pitaevskii2003bose}. We find that the condensate fraction on a fuzzy sphere is larger than in case of a commutative theory. Moreover,  $T_\text{BEC}$  increases with an increase in the non-commutativity parameter. Similar to BEC in an ideal Bose gas, fuzziness enhances BEC also in an interacting system. 

Sections \ref{Sect:SF-density} and \ref{Sect:Fuzzy-vortex} study superfluidity on a fuzzy sphere.  
In Section \ref{Sect:SF-density}, the superfluid density is calculated in the Bogoliubov approximation from the current correlation function. As in the previous sections, the most relevant results are obtained for a fuzzy sphere with a large enough area to justify a thermodynamic limit.  In this regime, the superfluid fraction is shown to be larger for higher values of the non-commutativity parameter, highlighting once more the propensity of fuzziness to increase order.   Furthermore, the  large-$R$ result  for the normal fluid density on a fuzzy sphere yields a behavior linear in $T$, which is compatible with Uemura's law~\cite{Uemura_PhysRevLett.62.2317}. Thus, the superfluid density at the critical temperature $T_c$ is given by $\rho_s(T_c)=K T_c$, where $K$ is some constant closely related to the non-commutativity parameter. This result contrasts with the commutative sphere limit ($N\to\infty$), where the normal fluid density is independent of the radius of the sphere and is identical to the one obtained for the planar case, behaving as $\sim T^3$ with the temperature~\cite{Fisher-Hohenberg_PhysRevB.37.4936}. This remarkable difference in the temperature-dependence of the normal fluid density highlights  that the fuzzy sphere has a distinct thermodynamic limit, in which Uemura's law is obtained solely as a consequence of non-commutativity.

Moving on to account for vortices in a fuzzy superfluid, we arrive at Section \ref{Sect:Fuzzy-vortex}. Since the fuzzy sphere considered in this work is a two-dimensional surface, it is natural to investigate the similarities with the BKT vortex-unbinding transition occurring in the planar case. Thus, Section  \ref{Sect:Fuzzy-vortex} has several subsections, where the first one briefly reviews the role of vortices in a planar geometry leading to the BKT transition. The second subsection showcases non-commutativity as an inherent feature of the collective coordinates (moduli) \cite{manton2004topological} of vortex centers, a result that follows from the analysis by Haldane and Wu \cite{Haldane-Wu_PhysRevLett.55.2887} in the case of a planar superfluid. The third subsection    
considers vortices on a commutative $S^2$ sphere, which remarkably have features in common with vortices on a plane. In particular, the critical point can be located via the same formula as in the planar case. 
This again demonstrates the equivalence between the $S^2$ and a plane in the thermodynamic limit, while pointing toward a more distinct behavior of vortices on a fuzzy sphere. 
 The fourth subsection extends the Haldane-Wu formalism to the case of $S^2$. The final subsection in Section~\ref{Sect:Fuzzy-vortex} provides a construction of vortices on a fuzzy sphere. This construction is more challenging compared to both the planar and $S^2$ cases, since the notion of a point vortex is not warranted due to the non-commutativity of space. Our conclusions and final remarks are presented in Section \ref{Sect:Conclusions}.

\section{Bose-Einstein condensation for an ideal Bose gas on a fuzzy sphere}
\label{Sect:BEC-fuzzy}

\subsection{The Lagrangian of an ideal Bose gas on a fuzzy sphere}

 As anticipated earlier, we introduce a complex Hermitian $N\times N$ matrix field $\psi$, which corresponds to a complex scalar boson in the commutative theory. Since in our case space is fuzzy, but not time, we construct the following action for the nonrelativistic free theory in imaginary time, 
\begin{eqnarray}
	\label{Eq:Ideal-Bose}
	S_0&=&\frac{4\pi R^2}{N}\int_{0}^\beta d\tau\left\{\vphantom{\frac{1}{R^2}}{\rm Tr}[\psi^\dagger(\partial_\tau-\mu)\psi]
	\right.\nonumber\\
	&+&\left.\frac{1}{2R^2}{\rm Tr}\left\{[L_i,\psi]^\dagger[L_i,\psi]\right\}\right\}. 
\end{eqnarray}

Let us consider the decomposition of the field into angular momentum modes, 
\begin{equation}
	\label{Eq:psi-expan}
	\psi=\sqrt{\frac{N}{4\pi R^2}}\sum_{l=0}^M\sum_{m=-l}^{l}\eta_{l m}T_{lm},
\end{equation}
where $M=N-1$, $\eta_{l m}$ are complex fields, and $T_{lm}$ are matrices corresponding to representations of $SU(2)$ \cite{Nair_PhysRevD.85.045021}. Taking into account the faact that the second term in the action can be rewritten as
\begin{equation}
	{\rm Tr}\left\{[L_i,\psi]^\dagger[L_i,\psi]\right\}={\rm Tr}\left\{\psi^\dagger[L_i,[L_i,\psi]]\right\}
	={\rm Tr}\left[\psi^\dagger L^2\psi\right],
\end{equation}
one obtains in terms of the field components $\eta_{l m}$, 
\begin{equation}
	\label{Eq:S0}
	S_0=\sum_{l=0}^M\sum_{m=-l}^{l}\int_{0}^\beta d\tau\eta_{l m}^*(\tau)\left[\partial_\tau-\mu+\frac{l(l+1)}{2R^2}\right]\eta_{l m}(\tau). 
\end{equation}
Within a path integral formalism, the action \eqref{Eq:S0} is reminiscent of a second-quantized formulation of the quantum rotor model, but where $l$ is subjected to a maximum value, $M$. Note from this comparison that we are not only employing a unit system where $\hbar=c=1$, but we are also setting the mass to unity, so the moment of inertia is simply $R^2$. Since $\eta_{l m}$ are bosonic fields, the path integral is solved with a periodic boundary condition, $\eta_{l m}(0)=\eta_{l m}(\beta)$, in the Matsubara time.    

\subsection{Bose-Einstein condensation}

The exact partition function obtained by integrating out $\eta_{l m}$ is given by,
\begin{equation}
	Z=\prod_{l=0}^{M}\left[1-e^{-\beta(\epsilon_l-\mu)}\right]^{-(2l+1)}, 
\end{equation}
where we have defined, $\varepsilon_l=\frac{l(l+1)}{2 R^2}$.
This immediately leads to the particle number, 
\begin{equation}
	 \mathcal{N}=\sum_{l=0}^M \frac{2 l+1}{e^{\beta(\varepsilon_l-\mu)}-1}, 
\end{equation}
 
 As usual in standard BEC, the $l=0$ mode becomes singular for $\mu=0$, signaling the onset of BEC in the thermodynamic limit. In order to analyze this regime properly, we consider the particle density, $n=\mathcal{N}/(4\pi R^2)$. The thermodynamic limit corresponds to taking both $\mathcal{N}$ and the area $4\pi R^2$ as infinitely large quantities, while keeping $n$ fixed. This allows for a finite condensate fraction, $n_0$ when $\mu=0$. Later on, when we consider the weakly interacting case, we will see that there are some subtleties involved regarding the interplay between the $M$ large regime  and the thermodynamic limit itself. 
 
 After separating the condensate part, $n_0$, we can use the fact that $R$ is large and approximately replace the sum over $l$ by an integration in the interval $[1,M]$ to evaluate the part $n-n_0$ of the particle density out of the condensate,    
	\begin{eqnarray}
	 n&\approx& n_0+\frac{1}{4 \pi R^2} \int_1^M d l \frac{2 l+1}{e^{\beta\varepsilon_l}-1},
	\end{eqnarray}
which yields, 	
\begin{equation}
	n=n_0+\frac{1}{2\pi\beta}\ln\left(\frac{1-e^{-\beta\varepsilon_M}}{1-e^{-\beta\varepsilon_1}}\right).
\end{equation}
Note that this way of defining a thermodynamic limit precludes the use of too small values of $M$: replacement of the sum over $l$ by an integral over $l$ is a good approximation only when $M$ is sufficiently large. 
Essentially the rationale here does not differ much from that for replacement of the sum over momenta by an integral over momenta more standard in calculations characterizing condensation.
A similar procedure is also applied to BEC in a trap \cite{pitaevskii2003bose}. The main difference in the case of a fuzzy sphere is the presence of the upper cutoff, $M$, which underlies the non-commutativity parameter.   

From the condition $n_0=0$ the critical temperature below which the system undergoes BEC is obtained as a transcendental equation, 
\begin{equation}
	\label{Eq:Tc-ideal-fuzzy}
	T_{\text{BEC}}=\frac{2 \pi n}{\ln\left(\frac{1-e^{-\frac{\varepsilon_M}{T_{\rm BEC}}}}{1-e^{-\frac{\varepsilon_1}{T_{\rm BEC}}}}\right)}. 
\end{equation} 
The dashed lines in Figure \ref{Fig:T-nR2-inter} show the dependence of $T_{\rm BEC}/n$ on $nR^2$ for different values of $M$ in an ideal Bose gas. We see from the figure that $T_{\rm BEC}/n$ grows as $M$ decreases for large enough $nR^2$, meaning that the BEC fraction becomes larger with increasing fuzziness, corresponding to decreasing $M$, if $nR^2$ is larger than $\sim 10^3$. Interestingly, for $nR^2$ smaller than 500 all the curves nearly overlap. Note, however, that $M$ cannot be arbitrarily small, something we have alluded to above. Even though by means of a sum over $l$ instead of performing an integral over $l$, it is technically possible to find a critical temperature, the results lose physical significance, since it is possible to find a nearly infinite value of $T_{\text{BEC}}$. This is an artifact related to an improper handling of the thermodynamic limit.

In the $M\to\infty$ regime the result of Eq. \eqref{Eq:Tc-ideal-fuzzy} reduces to the one of a commutative (non-fuzzy) sphere \cite{Tononi2019},
\begin{equation}
	\label{Eq:Tc-ideal-non-fuzzy}
	T_{\rm BEC}^{S^2}=\lim_{M\to\infty}T_{\rm BEC}=\frac{2 \pi n}{\ln \left(1-e^{-\frac{1}{R^2 T_{\text{BEC}}^{S^2}}}\right)}.
\end{equation} 
Note that $T_{\rm BEC}^{S^2}$ vanishes for $R\to\infty$. This is to be expected, as an infinite radius corresponds to the limit of a flat two-dimensional system, where no BEC occurs at any finite temperature \cite{Hohenberg_PhysRev.158.383}.

\section{Bose-Einstein condensation in a weakly interacting Bose gas on a fuzzy sphere}
\label{sec:intBEC}

In this section we will treat the weakly interacting theory in the Bogoliubov approximation. The action to be considered is, 
\begin{equation}
	S=S_0+\frac{g}{2}\frac{4\pi R^2}{N}{\rm Tr}\left[(\psi^\dagger\psi)^2\right]. 
\end{equation}

In the Bogoliubov approximation we decompose the bosonic field as a sum, $\psi=\sqrt{N}\psi_0+\eta$, where $\psi_0$ is a uniform condensate and $\eta$ is a fluctuation term. Then, assuming the fluctuations are small, we only keep terms up to second order in $\eta$.   The action takes the form, $S=S_{\psi_0}+S_\eta$, where the condensate part is $S_{\psi_0}= 4\pi R^2\beta\left(-\mu|\psi_0|^2+\frac{g}{2}|\psi_0|^4\right)$ and $S_\eta$  is the fluctuations contribution. The latter is obtained by  performing the Fourier transformation in time and using the angular momentum representation~\eqref{Eq:psi-expan} for the fluctuation fields, 
\begin{eqnarray}
	S_\eta&=&\sum_{n} \sum_{l=1}^M \sum_{m=-l}^{+l}\eta_{l m}^{*}\left(-i \omega_n+\varepsilon_l-\mu+2 g\left|\psi_0\right|^2\right) \eta_{l m}\nonumber\\&+&\sum_{n} \sum_{l=1}^M \sum_{m=-l}^{+l}\frac{g}{2}\left[\psi_0^2\left(\eta_{l m}^{*}\right)^2+\left(\psi_0^{*}\right)^2 \eta_{l m}^2\right],
\end{eqnarray}
where $\omega_n=2\pi n/\beta$ is the bosonic Matsubara frequency. 
This action is more conveniently written in matrix form as, 
\begin{widetext}
 \begin{equation}
		S_\eta=\frac{1}{2} \sum_{n} \sum_{l=1}^M \sum_{m=-l}^{+l}\left(\eta_{l m}^*\left(\omega_n\right) \quad \eta_{l,-m}\left(-\omega_n\right)\right) \hat{M}_{lm}(\omega_n)\binom{\eta_{l m}\left(\omega_m\right)}{\eta_{l,-m}^*\left(-\omega_n\right)},
\end{equation}
where we have defined the matrix, 
\begin{equation}
	\label{eq:mlm}
	\hat{M}_{lm}(\omega_n)=
	\left[
	\begin{array}{cc}
		-i \omega_n+\varepsilon_l-\mu+2 g\left|\psi_0\right|^2 & g\psi_0^2\\
		\noalign{\medskip}
	g(\psi_0^*)^2 & i \omega_n+\varepsilon_l-\mu+2 g\left|\psi_0\right|^2
	\end{array}
	\right].
\end{equation}
\end{widetext}
Integrating out the Gaussian fluctuations, we obtain the  partition function $Z$, which allows us to calculate the free energy density, $\mathcal{F}=-\frac{1}{4\pi R^2 \beta}\ln Z$. Specifically, 
\begin{eqnarray}
	\mathcal{F}&=&-\mu|\psi_0|^2+\frac{g}{2}|\psi_0|^2
	\nonumber\\
	&+&\frac{1}{8 \pi R^2 \beta} \sum_{n} \sum_{l=1}^M \sum_{m=-l}^{+l} \ln \operatorname{det} \hat{M}_{lm}.
\end{eqnarray}
The particle number density, $n=-\frac{\partial \mathcal{F}}{\partial \mu}$, now becomes, 
\begin{equation}
	\label{Eq:n-before-sum}
	n=\left|\psi_0\right|^2+\frac{1}{4 \pi R^2 \beta}\sum_{n} \sum_{l=1}^M (2l+1)\frac{i \omega_n+\varepsilon_l+ g\left|\psi_0\right|^2}{\omega_n^2+E_l^2},
\end{equation}
where we have applied the minimization condition, $\frac{\partial S_{ \psi_0}}{\partial |\psi_0|^2}=0$ to obtain the relation, $\mu = g|\psi_0|^2$, and, 
\begin{equation}
	E_l=\sqrt{\varepsilon_l\left(\varepsilon_l+2 g  |\psi_0|^2\right)},
\end{equation}
is the Bogoliubov spectrum for this system. 

 As in the previous Section, we approximate the sum $\sum_{l=1}^M$ by an integral $\int_1^Mdl$. We proceed by first evaluating the term containing $i \omega_n$. In this case, it is more useful to start by summing over the Matsubara frequencies and then integrating over $l$. Since 
\begin{equation}
	\frac{1}{\beta}\sum_{n} \frac{i \omega_n (2l+1)}{\omega_n^2+E_l^2}=-\frac{1}{2}(2l+1),
\end{equation}
we get 
\begin{equation}
	\frac{1}{4 \pi R^2 \beta} \int_{l=1}^M dl	\sum_{n} \frac{i \omega_n (2l+1)}{\omega_n^2+E_l^2}=-\frac{1}{8\pi R^2} (M^2+M-2).
\end{equation}
Evaluating the other terms in Eq.~\eqref{Eq:n-before-sum}, it is more convenient to first integrate over $l$ and then sum over $n$,
\begin{eqnarray}
	&&\frac{1}{4 \pi R^2 \beta}\sum_{n} \sum_{l=1}^M \frac{2l+1}{\omega_n^2+E_l^2}\left[\frac{l(l+1)}{2R^2}+ g\left|\psi_0\right|^2\right]
	\nonumber\\
	&=&\frac{1}{4 \pi R^2 \beta} \sum_n \ln \left(\omega_n^2+\frac{x^2}{4 R^4}+g\left|\psi_0\right|^2 \frac{x}{R^2}\right)\Biggr|_2 ^{M(M+1)}.
\end{eqnarray}
Since
 \begin{equation}
	\sum_n \ln \left(\omega_n^2+\omega^2\right)=\beta \omega+2 \ln \left(1-e^{-\beta \omega}\right)-\ln 2,
\end{equation}
we obtain the particle number density,
\begin{equation}
	\label{Eq:total-particle-density}
	n=n_0 +\delta n_0 + \delta n_{\text{th}},
\end{equation}
where the temperature-independent contribution is 
\begin{eqnarray}
		\label{Eq:depletionT0}
	\delta n_0  &=& \frac{1}{8 \pi R^2} \sqrt{M^2(M+1)^2+4 M(M+1) g n R^2}
	\nonumber\\
	&-&\frac{1}{4 \pi R^2} \sqrt{1+2 g n R^2}-\frac{1}{8\pi R^2} (M^2+M-2),
	\nonumber\\
\end{eqnarray}
and the particle number density fluctuations following from thermal effects are given by
\begin{eqnarray}
		\label{Eq:depletionTfinite}
	\delta n_{\text{th}} &=& \frac{1}{2 \pi \beta} \ln \left(1-e^{-\frac{\beta}{2 R^2} \sqrt{M^2(M+1)^2+4 M(M+1) g n R^2}}\right)
	\nonumber\\
	&-&\frac{1}{2 \pi \beta} \ln \left(1-e^{-\frac{\beta}{ R^2} \sqrt{1+2 g n R^2}}\right).
\end{eqnarray}
In the above equations for the particle number density fluctuations we have  replaced $|\psi_0|^2$ by $n$, since the error committed with this substitution is of higher order in $g$. 

\begin{figure}[h]
	\includegraphics[width=0.99\linewidth]{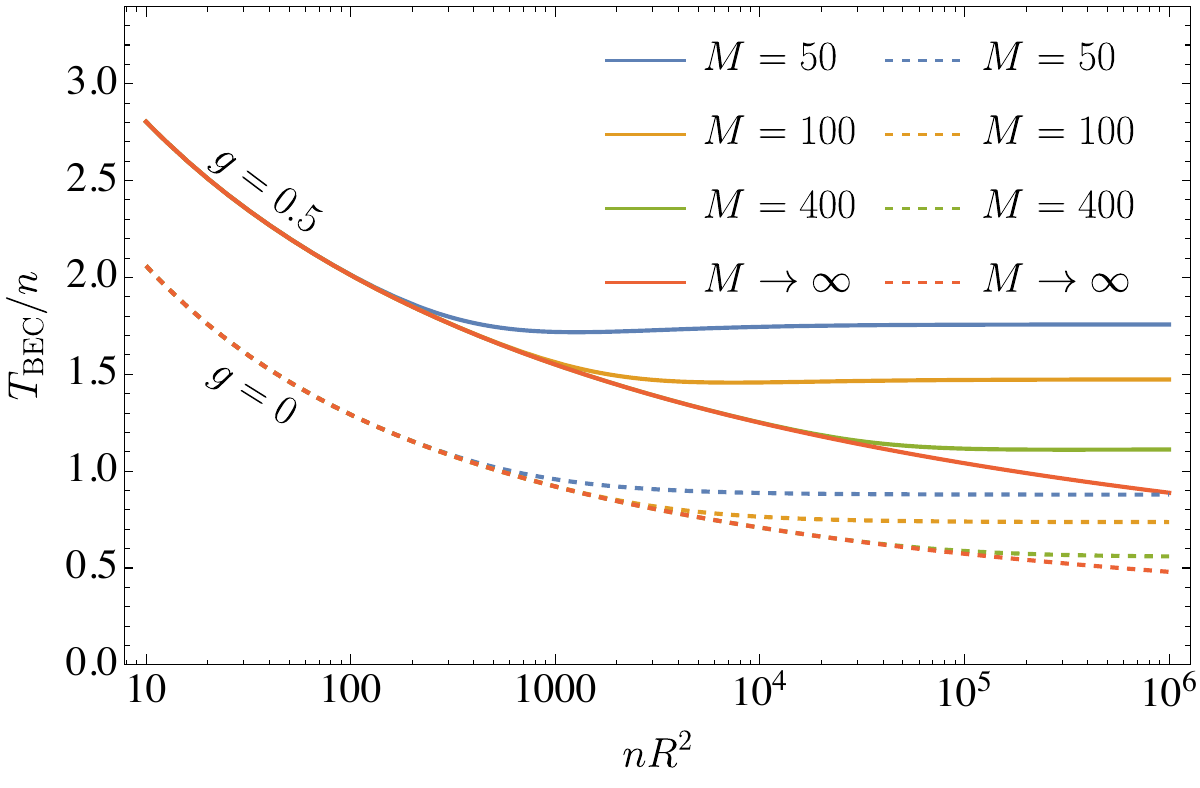}
	\hfill
	\caption{Behavior of $T_{\rm BEC}/n$ as a function of $nR^2$ for different values of $M$. The solid lines correspond to the weakly interacting Bose gas (specifically, here we take $g=0.5$), while the dashed lines correspond to the ideal case [see Eq.~\eqref{Eq:Tc-ideal-fuzzy}]. In both cases, for a fuzzy sphere of a given radius the critical temperature $T_{\rm BEC}$  grows as the non-commutativity parameter is increased. The influence of non-commutativity is most noticeable in the large $R$ limit, where the calculations are most applicable. }
	\label{Fig:T-nR2-inter}
\end{figure}

\begin{figure}[h]
	\includegraphics[width=0.99\linewidth]{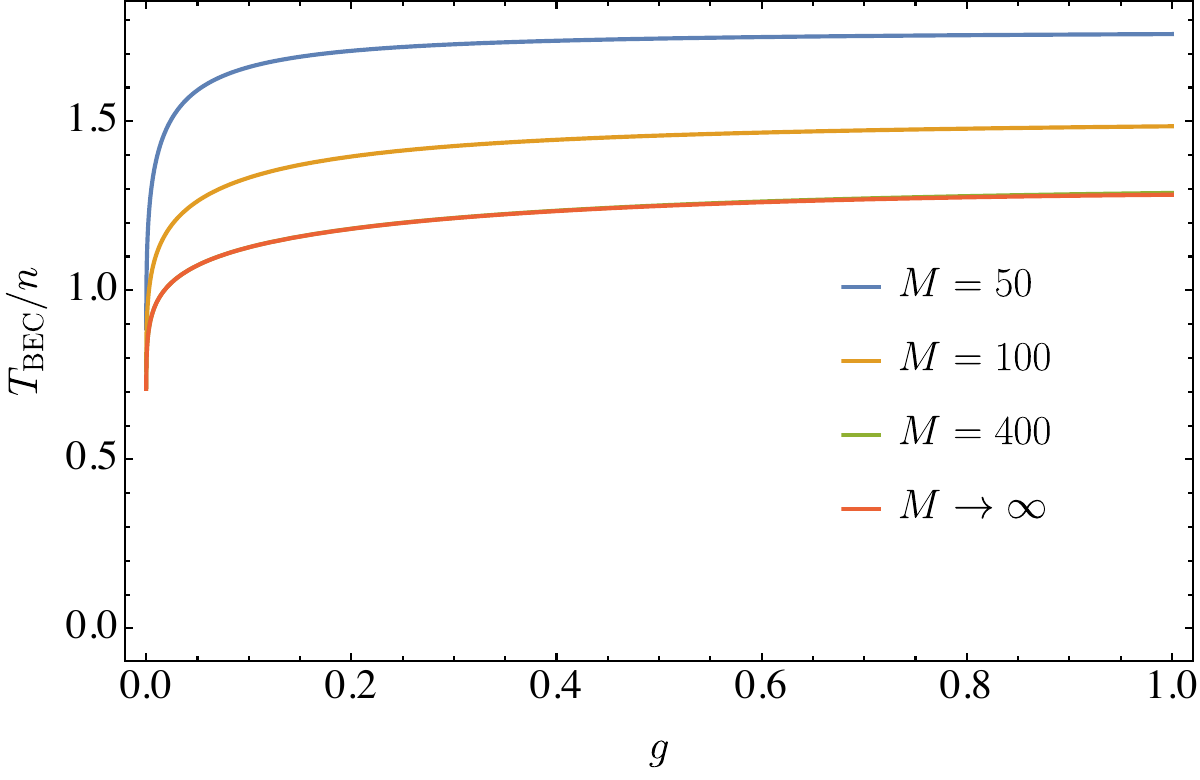}
	\hfill
	\caption{Dependence of $T_{\rm BEC}/n$ on the interaction coupling $g$ for different values of $M$, where $nR^2$ is set to be equal to $10^4$. Similarly to the behavior captured in Fig.~\ref{Fig:T-nR2-inter}, the increase in the non-commutativity parameter (decrease in $M$) leads to higher values of  $T_{\rm BEC}$. Already at $M=400$, the fuzzy result has basically the same dependence on $g$ as the commutative limit $M\to \infty$.}
	\label{Fig:T-ginter}
\end{figure}

It is possible to obtain an analytical self-consistent equation for $T_\text{BEC}$ in the weakly interacting Bose gas (similar to the one presented for an ideal case in Eq.~\eqref{Eq:Tc-ideal-fuzzy}) from the particle number density in Eq.~\eqref{Eq:total-particle-density}, however, the expression is lengthy and not particularly illuminating. 
Thus, we simply provide the plot depicting this solution in Figs.~\ref{Fig:T-nR2-inter} and~\ref{Fig:T-ginter}.  Note that the critical temperature  $T_\text{BEC}$  for the weakly interacting case is larger than the one for the ideal Bose gas. This behavior is reminiscent of the upwards shift in the critical temperature of the weakly interacting gas compared to the one in the ideal case in $\mathbb{R}^3$ \cite{Kastening2003, Kleinert-Schmidt-Pelster-2004}. 

There are some interesting aspects of the interaction-induced depletion of the condensate at $T=0$,
\begin{eqnarray}
	\label{Eq:depletion-fuzzy}
	n&=&n_0+\frac{M(M+1)}{8 \pi R^2} \sqrt{1+\frac{4 gn R^2}{M(M+1)}}
	\nonumber\\
	&-&\frac{\sqrt{1+2 gn R^2}}{4 \pi R^2}-\frac{M^2+M-2}{8 \pi R^2}.
\end{eqnarray}
First, it reduces in the $M\to\infty$ limit to the result obtained in Ref. \cite{Tononi2019} for an interacting Bose gas on a non-fuzzy sphere,
\begin{equation}
	\label{Eq:depletion}
	n=n_0+\frac{gn}{4\pi}+\frac{1-\sqrt{1+2gnR^2}}{4\pi R^2}. 
\end{equation}
Second, the resulting expression reduces in turn to the result for the $\mathbb{R}^2$ plane \cite{Schick_PhysRevA.3.1067} in the limit $R\to\infty$. 
Note that the limits $M\to\infty$ and $R\to\infty$ do not commute. In order to recover the result of Ref. \cite{Schick_PhysRevA.3.1067}, it is crucial to first remove the fuzziness of the sphere by letting $M\to\infty$. 

This  prompts an important remark about the defintion of thermodynamic limit on a fuzzy sphere. The calculations we have done so far assume that $M$ is large in such a way that the sum over $l$ can be replaced by an integral. As we have seen, there are essentially two ways of making $M$ large. Since $M$ is an integer, there is the absolute way, which is simply taking $M\to\infty$, which completely removes the non-commutativity of space. The other approach is to keep the non-commutativity parameter, $2R/N=2R/(M+1)$, fixed and finite, while assuming that {\it both} $M$ and $R$ are large. However, to go to thermodynamic limit in either of these two cases (commutative and non-commutative), one has to additionally send the particle number and the sphere area to infinity, while keeping their ratio, the particle density, fixed. The difference between the approach where the non-commutativity parameter remains finite even for $R\to \infty$ and $M\to \infty$ and the commutative approach where $M$ is sent to infinity so that the non-commutativity parameter vanishes, can be illustrated by Eqs. \eqref{Eq:depletion-fuzzy} and \eqref{Eq:depletion}. If one simply takes the thermodynamic limit, $R\to\infty$, in Eq.~ \eqref{Eq:depletion}, the last term vanishes and there is no way to distinguish this commutative sphere result from the depletion on the plane. On the other hand, the thermodynamic limit for a fuzzy sphere, which keeps the non-commutativity parameter fixed (although $R\to \infty$ and $M\to \infty$), when applied to Eq.~\eqref{Eq:depletion-fuzzy}, retains the terms stemming from fuzziness and is distinct from the planar case.

\section{Superfluid density on a fuzzy sphere}
\label{Sect:SF-density}

In contrast to the condensate density, the superfluid density, $\rho_s$,  does not follow from the expectation value of the particle number operator. It is given by the response of the system to a Galilei boost, to which normal fluid does not react in view of its Galilei invariance \cite{pitaevskii2003bose,Weichmann_PhysRevB.38.8739}. This response is very generally captured by a Kubo-type formula, $\rho_s=\rho-\rho_n$, where $\rho=\left\langle | \psi|^2\right\rangle$ is the total fluid density\footnote{In this instance, the fluid density is actually the same as the particle density, $n$, used in the previous section. Since the fluid mass density is given by $\rho= mn$, where $m$ is the mass of the bosonic particle, and we set $m=1$, $\rho$ is actually identical to $n$ in our case. However, in this section we will use the symbol $\rho$ to follow the more standard notation for discussing superfluid density. }  
and the normal fluid density, $\rho_n$, is given by \cite{Fisher_PhysRevLett.64.587}, 
\begin{equation}
	\rho_n = \frac{1}{D}\frac{1}{V\beta} \int_0^\beta d \tau \int d^Dr \int_0^\beta d \tau' \int d^Dr'C(\tau,\vec{r};\tau',\vec{r}'),
\end{equation}
where $V$ is the volume and, 
\begin{equation}
	C(\tau,\vec{r};\tau',\vec{r}')=\left\langle \vec{j}(\tau, \vec{r})\cdot \vec{j}(\tau', \vec{r}')\right\rangle-\left\langle \vec{j}(\tau, \vec{r})\right\rangle\cdot \left\langle\vec{j}(\tau', \vec{r}')\right\rangle,
\end{equation}
is the current correlator, which in the commutative case is associated to the $U(1)$ current, 
\begin{equation}
	\label{Eq:current}
	\vec{j} = \frac{i}{2}\left(\psi\nablab\psi^*-\psi^*\nablab\psi\right).
\end{equation}

Now we recall that we have mentioned in the Introduction that on a sphere the momentum operator should be considered as $\vec{L}/R$. This leads us to consider for $S^2$ instead, 
\begin{equation}
	\label{Eq:non-fuzzy-current}
	\vec{j} = \frac{1}{2R}\left(\psi\vec{L}\psi^*-\psi^*\vec{L}\psi\right)=\frac{i}{2}\hat{\vec{r}}\times(\psi\nablab\psi^*-\psi^*\nablab\psi).
\end{equation}

On the other hand, in the case of a fuzzy sphere, each component of the above current becomes a matrix of the form, 
\begin{eqnarray}
	\label{Eq:Fuzzy-current}
	j_\alpha&=&\frac{i}{2R}\left[\psi(\mathcal{L}_\alpha\psi)^\dagger-\psi^\dagger\mathcal{L}_\alpha\psi\right]
	\nonumber\\
	&=&-\frac{1}{2R}\left\{\psi^\dagger[L_\alpha,\psi]-\psi[L_\alpha,\psi]^\dagger\right\}, 
\end{eqnarray}
where we have employed the definition of the Lie derivative from the Introduction. 

Thus, a proper generalization of the Galilei Kubo response in the case of a fuzzy sphere is, 
\begin{equation}
	\rho_n=\frac{1}{2\beta}\int_{0}^{\beta}d\tau d\tau'{\rm Tr}\left[\langle j_\alpha(\tau)j_\alpha(\tau')\rangle-\langle j_\alpha(\tau)\rangle\langle j_\alpha(\tau')\rangle\right].
\end{equation}

We will consider the analogue of the hydrodynamic regime for a superfluid on a fuzzy sphere, meaning that we will employ the linear approximation for the energy of Bogoliubov quasi-particles,
\begin{equation}
	E_l^2\approx 2g\varepsilon_l\left|\psi_0\right|^2,
\end{equation}
which is valid for large $R$. As in the previous Section, we approximate once more, by taking $|\psi_0|^2=n$, and write $E_l^2=v^2l(l+1)/R^2$, where $v^2=g|\psi_0|^2=gn$. After some algebra involving Wick contractions, we obtain,  
\begin{equation}
	\label{Eq:rho-n}
	\rho_n=\frac{T}{2}\sum_{n=-\infty}^{\infty}\sum_{l=0}^{M}(2l+1)\frac{\frac{l(l+1)}{R^2}\left[-\omega_n^2+\frac{v^2l(l+1)}{R^2}\right]}{\left[\omega_n^2+\frac{v^2l(l+1)}{R^2}\right]^2}.
\end{equation}
Further calculations detailed in Appendix \ref{App:rhon} finally yield, 
\begin{eqnarray}
	\label{Eq:rhon-final}
	4\pi \rho_n&=&
	\frac{[M(M+1)]^{3/2}}{2vR^3}\left[1-\coth\left(\frac{v\sqrt{M(M+1)}}{2RT}\right)\right]
	\nonumber\\
	&+&\frac{3T}{v^2R^2}M(M+1)\ln\left(1-e^{-v\sqrt{M(M+1)}/(RT)}\right)
	\nonumber\\
	&-&\frac{6T^2}{v^3R}\sqrt{M(M+1)}{\rm Li}_2\left(e^{-v\sqrt{M(M+1)}/(RT)}\right)
	\nonumber\\
	&-&\frac{6T^3}{v^4}{\rm Li}_3\left(e^{-v\sqrt{M(M+1)}/(RT)}\right)+\frac{6T^3}{v^4}\zeta(3).
\end{eqnarray}

\begin{figure}[h]
	\includegraphics[width=0.99\linewidth]{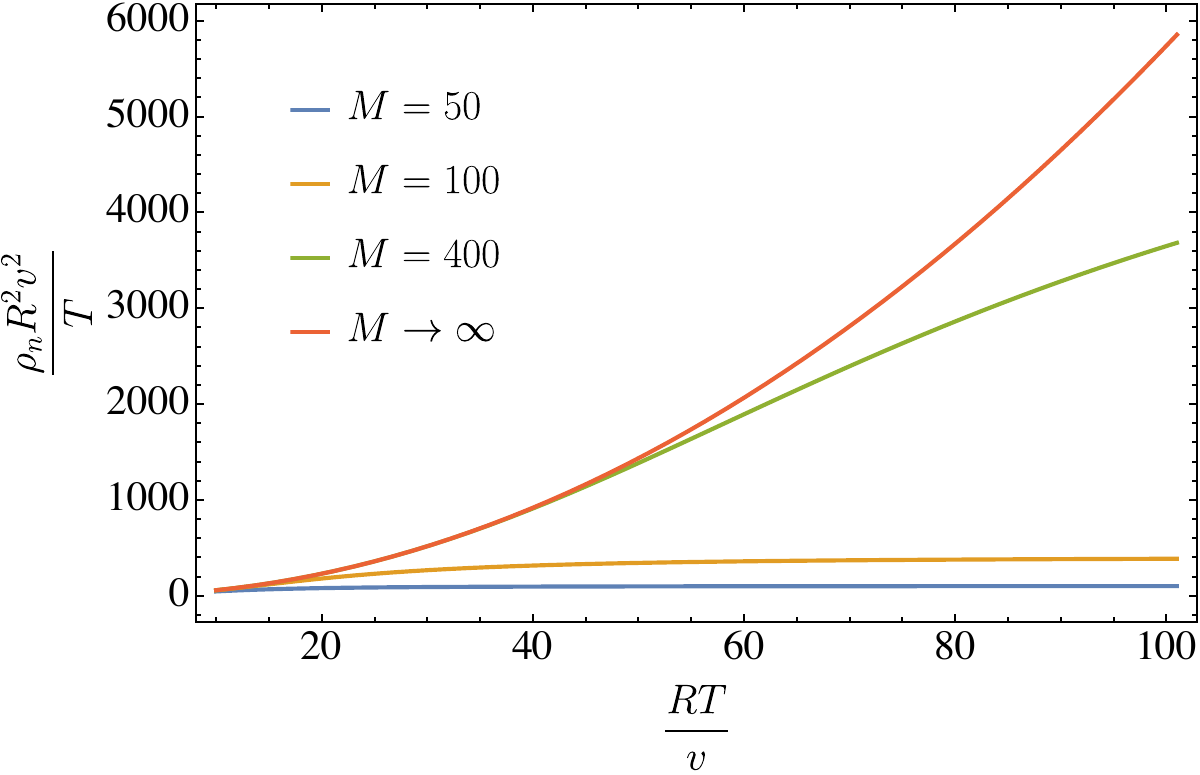}
	\hfill
	\caption{The normal fluid density $\rho_n$ as a function of the radius $R$ for various values of $M$ [see Eq.~\eqref{Eq:rhon-final}]. The normal fraction decreases for higher values of the non-commutativity parameter.}
	\label{Fig:normal-density}
\end{figure}

Figure \ref{Fig:normal-density} shows the behavior of $\rho_n$ as a function of $R$ for different values of $M$. We notice that the higher the fuzziness (i.e., by decreasing $M$), the higher the superfluid fraction will be. In view of this, it is interesting to consider the leading large $R$ behavior for a {\it finite} $M$. The result is, 
\begin{equation}
	\label{Eq:rhon-linear-T}
	\rho_n\approx \frac{M(M+1)T}{8\pi v^2R^2}.
\end{equation}
From this expression we clearly see that the critical temperature grows as $M$ decreases. 

On the other hand, for $M\to\infty$ Eq. \eqref{Eq:rhon-final} yields, 
\begin{equation}
	\label{Eq:rhon-cubic-T}
	\rho_n\underset{M\to\infty}{=}\frac{3T^3}{2\pi v^4}\zeta(3),
\end{equation}
which has exactly the same form as the result obtained in the long wavelength regime for the homogeneous case in $\mathbb{R}^2$ \cite{Fisher-Hohenberg_PhysRevB.37.4936}. Note that in contrast with the condensate density, the $M\to\infty$ limit is automatically yielding a result independent of $R$, so no $R\to\infty$ limit is required to obtain the planar case. This happens here because $R$ was considered large from the very beginning, since we assumed a linear approximation for the Bogoliubov spectrum, thus mimicking the long-wavelength limit underlying the hydrodynamic regime. Furthermore, recall that the thermodynamic limit itself requires taking the area of the fuzzy sphere, and thus $R$, to infinity, keeping the particle density fixed. Hence, a result for large $R$ which is indistinguishable from a planar one needs to account for the fuzziness of the sphere in such a way that $M$ is large with $R/M$ fixed. This allows us to obtain a robust critical behavior that is unaffected by the vanishing of the curvature of the commutative sphere. We have seen this subtlety before when considering the different ways of taking $M$ large in our discussion of the depletion of the condensate.   

Interestingly, the analysis of the limit cases in Eqs.~\eqref{Eq:rhon-linear-T} and \eqref{Eq:rhon-cubic-T} reveals two different power behaviors with the temperature that originate from the expression for the normal fluid density on a fuzzy sphere. The linear in $T$ behavior in the leading large $R$ regime of Eq. \eqref{Eq:rhon-linear-T} is reminiscent of  Uemura's law \cite{Uemura_PhysRevLett.62.2317} in cuprate high-$T_c$ superconductors. Indeed, it follows from Eq. \eqref{Eq:rhon-linear-T} that, 
\begin{equation}
	\label{Eq:Fuzzy-Uemura}
	\rho_s(T=0)=\frac{M(M+1)T_c}{8\pi v^2R^2}. 
\end{equation}
The non-fuzzy regime in Eq.~\eqref{Eq:rhon-cubic-T}, on the other hand, reproduces the planar Bogoliubov behavior, which is cubic in the temperature \cite{Fisher-Hohenberg_PhysRevB.37.4936}.

\section{Vortices on the fuzzy sphere}
\label{Sect:Fuzzy-vortex}

\subsection{Review of the planar case}

In any study of vortices in a $U(1)$ field theory the phase of the complex order parameter plays the most prominent role, especially in the two-dimensional case, as the BKT theory shows \cite{kosterlitz1973ordering,Nelson-Kosterlitz_PhysRevLett.39.1201}. It is typical in this case to consider a phase-only order parameter field, $\psi=\sqrt{\rho_0}e^{i\Theta}$, with the amplitude $\rho_0$ kept uniform. In $\mathbb{R}^2$ this yields an effective Hamiltonian, 
\begin{equation}
	\label{Eq:H-superfluid}
	H=\int d^2r \frac{\rho_0}{2}(\nablab\Theta)^2.
\end{equation}
From Eq. \eqref{Eq:current} we obtain, $\vec{j}=\rho_0\nablab\Theta$, which leads to the identification of $\vec{v}_s=\nablab\Theta$ as the superfluid velocity. We briefly review this case first, before moving on to consider both the sphere and its fuzzy counterpart. 

In $\mathbb{R}^2$ the superfluid velocity is decomposed into a longitudinal and a transverse component \cite{Nelson-Kosterlitz_PhysRevLett.39.1201}, 
\begin{equation}
	\label{Eq:vs}
	\vec{v}_s=\nablab\vartheta+\hat{\vec{z}}\times\nablab\Phi,
\end{equation}
with $\Phi$ satisfying the two-dimensional Poisson equation, 
\begin{equation}
	\label{Eq:Laplace-Stream}
	-\nabla^2\Phi=2\pi n_v(\vec{r}),
\end{equation}
where, 
\begin{equation}
	n_v(\vec{r})=\sum_aq_a\delta^2(\vec{r}-\vec{r}_a),
\end{equation}
is the point vortex density, which features integers $q_a$ satisfying the condition, $\sum_a q_a=0$. The second term in Eq. \eqref{Eq:vs} causes $\nablab\times\vec{v}_s\neq 0$ and corresponds to the vorticity term \cite{Nelson-Kosterlitz_PhysRevLett.39.1201}. The function $\Phi$ is sometimes called the stream function. 

The solution of Eq. \eqref{Eq:Laplace-Stream} is straightforward, and leads to the vortex or transverse contribution to the superfluid density,   
\begin{equation}
	\label{Eq:Transverse-vs}
	\vec{v}_s^T(\vec{r})=\frac{1}{2\pi}\sum_bq_b\frac{\hat{\vec{z}}\times(\vec{r}-\gammab^{(b)})}{|\vec{r}-\gammab^{(b)}|^2}. 
\end{equation}
Furthermore, this also yields directly from the Hamiltonian \eqref{Eq:H-superfluid} the vortex Hamiltonian, 
\begin{equation}
	\label{Eq:H-vortex-plane}
	H_v=-\frac{\rho_0}{2\pi}\sum_{a,b}q_aq_b\ln\left(\frac{|\gammab^{(a)}-\gammab^{(b)}|}{\xi_c}\right)+E_c\sum q_a^2,
\end{equation}
where $\xi_c$ is the length scale associate to the vortex core energy, $E_c$.  

On $\mathbb{R}^2$ the most accurate criterion for a BKT transition follows from the renormalization group analysis by Nelson and Kosterlitz \cite{Nelson-Kosterlitz_PhysRevLett.39.1201}, which predicts a universal jump in $\rho_s(T)/T$, as $T$ approaches $T_c$ from below, 
\begin{equation}
	\label{Eq:BKT-transition}
	\lim_{T\to T_c^-}\frac{\rho_s(T)}{T}=\frac{2}{\pi}. 
\end{equation}
Thus, the phase transition is a sharp one in the planar case.  

\subsection{Non-commutative behavior from vortices on the plane}

Setting $\vec{r}=\gammab^{(a)}$ in Eq. \eqref{Eq:Transverse-vs} yields the superfluid velocity of the $a$-th vortex center in the superfluid. Thus, we can describe the dynamics of vortices by demanding that $\vec{v}_s^T(\gammab^{(a)})=d\gammab^{(a)}/dt$. This yields, 
\begin{equation}
	\label{Eq:Vortex-dynamics}
	\frac{d\gammab^{(a)}}{dt}=\frac{1}{2\pi}\sum_{b\neq a}q_b\frac{\hat{\vec{z}}\times(\gammab^{(a)}-\gammab^{(b)})}{|\gammab^{(a)}-\gammab^{(b)}|^2}.
\end{equation} 
Haldane and Wu \cite{Haldane-Wu_PhysRevLett.55.2887} showed that Eq. \eqref{Eq:Vortex-dynamics} naturally follows from the Hamiltonian \eqref{Eq:H-vortex-plane}, provided $\gammab^{(a)}$ is promoted to be a quantum mechanical position operator satisfying the commutation relation, 
\begin{equation}
	\label{Eq:Vortex-position-commut}
	[\gamma_i^{(a)},\gamma_j^{(a)}]=i\frac{q_a}{2\pi\rho_0}\epsilon_{ij},
\end{equation}
where $\epsilon_{ij}$ is the two-dimensional Levi-Civita symbol.  
The position operators for different vortex centers commute. Hence, we can argue that in principle quantum vortices naturally induce spatial non-commutativity. Note, however, that this type of reasoning states that the vortex collective coordinates (moduli) \cite{manton2004topological}, upon quantization, do not commute. While this can in principle be used to motivate the non-commutative geometry of physical systems, in the example above it is not the bosonic field that is being fuzzified. 

\subsection{Vortices on $S^2$}

On a non-fuzzy (commutative) sphere $S^2$ we can use Eq. \eqref{Eq:non-fuzzy-current} to obtain, 
\begin{equation}
	\label{Eq:jS2}
	\vec{j}=\rho_0\hat{\vec{r}}\times\nablab\Theta,
\end{equation}  
 with the understanding that the gradient is taken with a fixed radial coordinate, $R$. This leads to a Hamiltonian like the one in Eq. \eqref{Eq:H-superfluid}, with the integration on the plane being replaced by an integration over the sphere $S^2$. We immediately note that Eq. \eqref{Eq:jS2} has a form similar to the second term of Eq. \eqref{Eq:vs}, but with the unit vector $\hat{\vec{z}}$ perpendicular to the plane replaced by the unit vector $\hat{\vec{r}}$ perpendicular to tangent plane of the sphere for a given point $\vec{r}$ on the sphere.  
 Separating the phases again into two contributions yields,  
\begin{equation}
	\label{Eq:vs-S2}
	\vec{v}_s=\hat{\vec{r}}\times(\nablab\vartheta+\nablab\Phi),
\end{equation}  
which should be regarded as a Hodge decomposition \cite{frankel2012geometry}. In this case, $\vartheta$ is a regular contribution. As a consequence, the so-called stream potential $\Phi$ satisfies a Poisson equation featuring a Laplace-Beltrami operator on $S^2$, 
\begin{equation}
		-\nabla_{S^2}^2\Phi=2\pi\sum_a q_a\delta^{2}(\Omegab-\Omegab_a),
\end{equation}
where $\Omegab$ is a unit vector and the neutrality condition $\sum_aq_a$ holds as before, just like in the planar case. The Green function of the Laplace-Beltrami operator is, 
\begin{equation}
	\label{Eq:Green-S2}
	G(\Omegab,\Omegab')=\sum_{l=1}^{\infty}\sum_{m=-l}^{l}\frac{Y_{lm}^*(\Omegab)Y_{lm}(\Omegab')}{l(l+1)}, 
\end{equation}
where the $l=0$ mode is removed as a regularization procedure. This is similar to the regularization done in the planar case, where the Green function at the origin is subtracted. Using the addition theorem for spherical harmonics, we rewrite Eq. \eqref{Eq:Green-S2} as, 
\begin{equation}
	G(\Omegab,\Omegab')=\frac{1}{4\pi} \sum_{l=1}^{\infty}\frac{2l+1}{l(l+1)}P_l(\Omegab\cdot\Omegab').
\end{equation} 
The final expression is obtained by making use of the identity \cite{Laplace-Beltrami-stuff}, 
\begin{equation}
	\sum_{l=1}^{\infty}\frac{2l+1}{l(l+1)}P_l(\Omegab\cdot\Omegab')=-1-\ln\left(\frac{1-\Omegab\cdot\Omegab'}{2}\right),
\end{equation} 
yielding, 
\begin{equation}
	\label{Eq:G-sphere}
	G(\Omegab,\Omegab')=-\frac{1}{4\pi}\ln\left(\frac{1-\Omegab\cdot\Omegab'}{2}\right)-\frac{1}{4\pi}.
\end{equation}
 At the end this yields the vortex contribution to the Hamiltonian, which has a form similar to that for the planar case, 
\begin{equation}
	\label{Eq:H-vortex-sphere}
	H_v=2\pi^2\rho_0\sum_{a,b}q_aq_bG(\Omegab_a,\Omegab_b)+E_c\sum q_a^2,
\end{equation}
where $E_c$ is the vortex core energy. Note that the constant term in Eq. \eqref{Eq:G-sphere} plays no role in the interaction between vortices in view of the neutrality condition, $\sum_a q_a=0$.

Strictly speaking, the criterion of Eq. \eqref{Eq:BKT-transition} for a BKT transition does not hold for $S^2$ in the case of an arbitrary area size of the sphere. At best a crossover behavior is expected, with a near BKT transition with logarithmic finite-size corrections when $R$ is large. However, overall a superfluid transition (though not a strict BKT one) is expected on the basis of the analysis of the superfluid density performed in Sect. \ref{Sect:SF-density} using the Bogoliubov approximation. Hence, in contrast to the planar case, for a sphere the Bogoliubov quasi-particles (phonon modes) provide the main drive for the superfluid transition. In spite of these considerations, it is intriguing that a short computation of the partition function for a vortex-antivortex pair yields, 
\begin{eqnarray}
	\label{Eq:Z-pair}
	Z_{\rm pair}\sim\int_{0}^{\pi}d\theta \frac{\sin\theta}{(1-\cos\theta)^{\pi\rho_0/(2T)}}\sim\frac{1}{1-\frac{\pi\rho_0}{2T}},
\end{eqnarray}   
where we assumed that one of the vortices is located at $\Omegab_0=(0,0,1)$, such that, $\Omegab\cdot\Omegab_0=\cos\theta$. Identifying the bare stiffness to the superfluid density at leading order, we obtain that $Z_{\rm pair}$ diverges for precisely the condition given in Eq. \eqref{Eq:BKT-transition} corresponding to the planar case. (More precisely, one should actually look at the pressure following from this partition function, as the transition in the planar case occurs when the pressure becomes negative, indicating an instability.) This result has been obtained before in Ref. \cite{Vortices-sphere_PhysRevD.43.1314} by computing the vortex dipole density, which is a calculation essentially similar to the one above. However, it would not be technically correct to identify this with a genuine critical point. Rather, it signals a crossover regime. Only when $R$ is large does this crossover  asymptotically approach the BKT regime. 

\subsection{Non-commutative behavior from vortices on $S^2$}

The Haldane-Wu formalism can easily be extended for the case of vortices on a sphere in way that is reminiscent from our discussion in the Introduction using the Dirac quantization of the electric charge-magnetic monopole pair (dyon). Accordingly, defining $\vec{n}=\vec{r}/R$, where $\vec{n}\in S^2$, it is well known that the problem of a {\it classical} spin, $S\vec{n}$, having a fixed angular momentum $S$, precessing around a magnetic field has the Poisson bracket, $\{n_i,n_j\}=S^{-1}\epsilon_{ijk}n_k$ \cite{STONE1989557}. For some generic vorticity $q\in\mathbb{Z}$, the spin is defined as, $S_q=\mathcal{N}|q|/2$, where $\mathcal{N}$ is the total number of particles on the sphere.
The role of the vortex collective coordinates on a sphere are here played by $x_i=Rn_i$, which we correspondingly associate to a quantum mechanical operator $\widehat{X}_i$. Thus, mapping the Poisson bracket to the operator commutator yields the commutation relation, 
\begin{equation}
	\label{Eq:Moduli-Commut}
	[\widehat X_i,\widehat X_j]=i\frac{R}{S_q}\epsilon_{ijk}\widehat X_k=i\frac{2R}{|q|\mathcal{N}}\epsilon_{ijk}\widehat X_k,
\end{equation}
which has precisely the same form as Eq. \eqref{Eq:Space-Commut}, but with $N$ replaced by $|q|\mathcal{N}$. It is remarkable that just as in the planar case discussed by Haldane and Wu \cite{Haldane-Wu_PhysRevLett.55.2887}, the vortex collective coordinates on a sphere are inherently non-commutative. Note, however, that we decided to be careful here and used hatted operator coordinates $\widehat{X}_i$ as a notation to emphasize the distinction between the collective coordinates and the actual operator coordinates, $X_i$, in the non-commutative space of the fuzzy sphere. Nevertheless, the discussion in this section serves to showcase yet another way to physically motivate the mathematical formalism underlying the fuzzy sphere. It is in fact intriguing that in the commutation relation \eqref{Eq:Moduli-Commut} the thermodynamic limit is naturally embedded in the non-commutative parameter, since $\mathcal{N}$ is the actual number of particles on the sphere.      

\subsection{Vortices on a fuzzy sphere}
\label{Subsect:Fuzzy-vortices}

We now turn to the study of vortices on a fuzzy sphere. So far we have seen that the collective coordinates of the
vortices  already define a non-commutative space, even in the planar case. The case of vortices on a sphere leads to vortex coordinates fulfilling a commutation relation of the form given in Eq. \eqref{Eq:Space-Commut} [see Eq. \eqref{Eq:Moduli-Commut}]. However, as we have pointed out, those coordinates are not the coordinates in general at which matrix bosonic fields are located. The purpose of this subsection is to fill this gap. We start by considering the bosonic field decomposition, $\psi=\sqrt{\rho_0}U$, where $U$ is a unitary matrix. Thus, Eq. \eqref{Eq:Fuzzy-current} yields, 
\begin{equation}
	j_\alpha=-\frac{\rho_0}{2R}\left\{U^\dagger[L_\alpha,U]-U[L_\alpha,U]^\dagger\right\}. 
\end{equation}
The sort of analytical approaches we employed in the previous two subsections for the commutative sphere do not work well in this case, since now we are dealing with matrices. As a consequence, the very notion of a point vortex cannot be defined in a strict sense. Rather, the delta function enforcing a point-like vortex density has to be replaced by an algebraic object. A suitable implementation is to use projectors from coherent states on the sphere \cite{Gazeau_2007}. By taking a projector $P_a=|\Omegab_a\rangle\langle\Omegab_a|$, where $|\Omegab_a\rangle$ is a $SU(2)$ coherent state we want to associate to a ``vortex", the analogue of the integration over a delta function yielding the unity is simply, ${\rm Tr}P_a=1$. Thus, we replace the delta function $\delta^2(\Omegab-\Omegab_a)$ by the operator, $\delta_a=N/(4\pi R^2)P_a$ (recall that $N=M+1$). This leads to a $\delta$ function-like operation on a given matrix $F$ playing the role of a function,  
\begin{equation}
	\frac{4\pi R^2}{N}{\rm Tr}\left(\frac{N}{4\pi R^2}P_a F\right)=\langle\Omegab_a|F|\Omegab_a\rangle. 
\end{equation}
In order to make this procedure more intuitive technically, we can make a closer parallel with the delta function on the sphere by writing something that is not a matrix and also represents a smeared coherent state point density (a fuzzy delta-like density). This is given by the action of the operator $\delta_a$ on a coherent state $|\Omegab\rangle$,
\begin{equation}
	\delta_a(\Omegab)=\left\langle\Omegab\left|\frac{N}{4\pi R^2}P_a\right|\Omegab\right\rangle=\frac{N}{4\pi R^2}|\langle\Omegab|\Omegab_a\rangle|^2. 
\end{equation}
From the theory of $SU(2)$ coherent states we obtain that \cite{gazeau2000coherent}, 
\begin{equation}
	|\langle\Omegab|\Omegab_a\rangle|^2=\left(\frac{1+\Omegab\cdot\Omegab_a}{2}\right)^M=\left[1-\frac{\left(\Omegab-\Omegab_a\right)^2}{4}\right]^M.
\end{equation}
Hence, 
\begin{equation}
		\delta_a(\Omegab)=\frac{(M+1)}{4\pi R^2}e^{M\ln\left[1-\frac{\left(\Omegab-\Omegab_a\right)^2}{4}\right]}. 
\end{equation}
Now we consider the large $M$ limit. Assuming that $|\Omegab-\Omegab_a|$ is small, we first obtain, 
\begin{equation}
	\delta_a(\Omegab)\approx\frac{(M+1)}{4\pi R^2}e^{-\frac{M}{4}\left(\Omegab-\Omegab_a\right)^2}, 
\end{equation}   
which in the $M\to\infty$ reduces to, 
\begin{equation}
		\delta_a(\Omegab)\sim \frac{1}{R^2}\delta^2(\Omegab-\Omegab_a).
\end{equation}
Therefore, we have obtained a definition of a delta function-like operator on a fuzzy sphere that reduces in the limit $M\to\infty$ to the delta function on $S^2$ divided by $R^2$. 

The important point to note in the above analysis is that for a fuzzy sphere the steps leading from Eq. \eqref{Eq:Green-S2} to Eq. \eqref{Eq:G-sphere} cannot be carried out for a finite $M$. Instead, the Green function is one where the sum over $l$ is cut off at $l_{\rm max}=M$, 
\begin{equation}
	\label{Eq:Green-S2-fuzzy}
	G_M(\Omegab,\Omegab')=\sum_{l=1}^{M}\sum_{m=-l}^{l}\frac{Y_{lm}^*(\Omegab)Y_{lm}(\Omegab')}{l(l+1)}. 
\end{equation}
The above Green function is used to solve the equation for the fuzzy stream potential, 
\begin{equation}
	[L_\alpha,[L_\alpha,\Phi]]=2\pi n_v^F,
\end{equation} 
where the fuzzy vortex density is given by, 
\begin{equation}
	\label{Eq:nv-fuzzy}
		n_v^F=\frac{N}{4\pi R^2}
		\left[\sum_a q_aP_a-\frac{1}{N}{\rm Tr}\left(\sum_a q_aP_a\right)\mathbf1\right].
\end{equation}
The role of the second term between brackets in Eq. \eqref{Eq:nv-fuzzy} is to explicitly enforce the neutrality constraint and guarantee that the mode $l=0$ is subtracted. 

The vortex Hamiltonian acquires a form similar to the one obtained in Eq.~\eqref{Eq:H-vortex-sphere} for a commutative sphere, but using the Green function \eqref{Eq:Green-S2-fuzzy} instead, 
\begin{equation}
	\label{Eq:H-vortex-fuzzy-sphere}
	H_v^F=2\pi^2\rho_0\sum_{a,b}q_aq_bG_M(\Omegab_a,\Omegab_b)+E_c^F\sum q_a^2.
\end{equation}
As a result, the would-be point vortex gets smeared with a characteristic fuzzy length, $\xi_F\sim R/\sqrt{N}$. As in the case of the commutative sphere, we do not expect any BKT transition to occur in the fuzzy sphere case. Here the logarithmic corrections due to finite size of the system (finite area of the sphere) are even stronger in view of the finite size of the would-be point vortices. 
Even when $R$ is very large, an actual transition should not be expected due to the presence of a finite $M$. However, as in the case of a commutative sphere, we showed in Sect. \ref{Sect:SF-density} that Bogoliubov quasi-particles can drive a genuine superfluid transition.  

A few remarks about the fuzzy  localization length $\xi_F$ are in order. First, it does not disappear in the thermodynamic scaling considered here. Indeed, since, 
\begin{equation}
	\xi_F^2\sim \frac{R^2}{N}
	=\frac{R}{2}\left(\frac{2R}{N}\right),
\end{equation}
keeping $2R/N$ fixed while $R,N\to\infty$ causes $\xi_F$ to grow as $\sqrt R$. At the same time, the maximal wave number $k_{\max}\sim M/R$ remains finite. It is this persistent spectral cutoff, associated with the non-commutative thermodynamic scaling, that changes the normal fluid density from the planar $T^3$ behavior to the linear in $T$ behavior found in Eq. \eqref{Eq:rhon-linear-T}, and consequently produces the Uemura-like relation in Eq. \eqref{Eq:Fuzzy-Uemura}.
The fuzzy length $\xi_F$ should not, however, be identified with the superconducting coherence length $\xi_{\rm sc}$ or directly with the size of a Cooper pair. In cuprate superconductors, $\xi_{\rm sc}$ is exceptionally short, whereas $\xi_F$ measures 
the minimal localization scale of a vortex on the fuzzy sphere. The analogy with the cuprates instead concerns the phase stiffness: the Uemura relation $T_c\propto \rho_s(0)$ suggests that the superconducting transition is controlled by the establishment of global phase coherence rather than directly by the pair formation scale. Thus, compact Cooper pairs and a short coherence length in the cuprates are compatible with Uemura scaling when the superfluid stiffness is small and phase fluctuations are important \cite{Emery-Kivelson}. In the fuzzy-sphere system, an analogous stiffness-controlled relation emerges through the geometry-induced truncation of collective modes, despite the very different microscopic meaning of $\xi_F$.

\section{Conclusions}
\label{Sect:Conclusions}

In this paper, we demonstrated that  a natural short-distance cutoff introduced by fuzziness leads to an enhancement of  BEC and superfluidity.  Indeed, both the ideal and interacting Bose systems studied here exhibit higher critical temperatures of BEC compared to ordinary (commutative) Bose gases.  This means that the higher the value of the non-commutativity parameter, the higher the critical temperature. 
However, effects of fuzziness are also prominent at $T=0$, as can be seen from the calculated depletion of BEC. There, the condensate fraction grows with the increase in the non-commutativity parameter.
Focusing on superfluidity, we calculated the normal fluid density~\eqref{Eq:rhon-final}, which possesses a much more complex form than  the one obtained in the planar commutative case. We found that the normal fluid fraction decreases with the increase of fuzziness. This effect is particularly remarkable in the large $R$ limit, where the superfluid density on the fuzzy sphere has a linear in $T$ behavior. 
This contrasts with the $T^3$ dependence one obtains in the commutative case and is reminiscent of Uemura's law in  high-$T_c$ cuprate superconductors. Uemura's law has been considered to require an exotic mechanism---ranging from non-Fermi liquid theories to holographic duality \cite{Nussinov-2002, Zaanen_Liu_Sun_Schalm_2015}---to be explained. However, here it arises as a direct consequence of the non-commutativity of space.

The non-commutativity enhancement of BEC and superfluidity can also be understood on the basis of fundamental thermodynamic arguments. This can be seen by comparing the thermal entropy for bosons on a fuzzy sphere and on the ordinary $S^2$ sphere. Already the ideal Bose gas can be used to illustrate this point. On a fuzzy sphere, the number of available one-particle modes is finite, $N_{\rm modes}=\sum_{l=0}^{M}(2l+1)=(M+1)^2=N^2$. Since the degeneracy is $g_l=2l+1$, the standard formula for the entropy in the microcanonical ensemble immediately yields for a fuzzy sphere, 
	\begin{equation}
		S_{\rm fuzzy}[\{n_l\}]=\sum_{l=0}^{M}\ln\binom{n_l+2l}{n_l},
	\end{equation}     
	where $n_l$ is the number of bosons occupying the level $l$. For BEC, we need the grand-canonical ensemble. In this case, for the ideal gas we have, 
	\begin{equation}
		S_{\rm fuzzy}=\sum_{l=0}^{M}(2l+1)s_l,
	\end{equation}
	where $s_l$ is the usual entropy of the level $l$ in the grand-canonical ensemble. Since the number of available states for the fuzzy sphere is finite, it is clear that the inequality, $S_{\rm fuzzy}<S_{S^2}$, holds, where $S_{S^2}$ denotes the entropy for the ideal Bose gas on a commutative sphere. Hence, non-interacting bosons on the fuzzy sphere have trivially less thermal entropy. 
	
	The obtained entropy inequality leads in turn to the expectation of a higher critical temperature, $T_{\rm BEC}$. A similar result holds for a weakly interacting Bose system. For the case of superfluidity, the breaking of Galilei invariance and the truncation of modes also reduces the number of thermally excited modes. This contributes to the current response, lowering in this way the normal fluid density and hence increasing the superfluid density. Therefore, fuzziness stabilizes both BEC and superfluidity by suppressing thermal phase space.

 The remark about the entropy  evokes the discussion about the definition of thermodynamic limit on a fuzzy sphere, which is  necessary to discuss the condensation. Here, we argued that in the case of a commutative sphere, $S^2$,  the thermodynamic limit involves essentially mapping the Bose system to the one with a planar ($\mathbb{R}^2$) geometry. 
 Since the Gaussian curvature of $S^2$  vanishes in the limit $R\to\infty$, locally the sphere $S^2$ is equivalent to a plane. 
 This is not the case for a fuzzy sphere, where one has to account for an additional scale, $N$, stemming from the size of the $N\times N$ matrices and appearing in the commutation relation for the space coordinates~ \eqref{Eq:Space-Commut}. On a fuzzy sphere, thermodynamic limit implies keeping the particle density and the non-commutativity parameter fixed, while $R$, $N$, and the particle number $	 \mathcal{N}$ are sent to infinity. This  preserves the influence of the sphere's curvature on  Bose-Einstein condensation and the superfluid phase transition.  
 Thus, unlike the commutative sphere, the fuzzy sphere in thermodynamic limit retains the information about its geometry and is distinct from the plane.  In fact, in view of the commutator ~\eqref{Eq:Space-Commut}, one can define on a fuzzy sphere a minimal elementary cell having area, $A=4\pi R^2/N$. Moreover, in Section \ref{Sect:Fuzzy-vortex} we have defined a fuzzy length, $\xi_F\sim R/\sqrt{N}$, that characterizes the smearing of what would otherwise be point vortices in the superfluid. Fuzziness endows the system with an additional geometric structure that survives the traditional thermodynamic limit, even if the (classical) Gaussian curvature still vanishes.

Having analyzed fuzzy superfluidity in the Bogoliubov quasi-particle approach, we continued to construct a way of introducing vortex defects into such systems. We adapted the approach by Haldane and Wu \cite{Haldane-Wu_PhysRevLett.55.2887}---showing that quantum vortices on a plane are intrinsically non-commutative---to the case where such vortices arise on a commutative sphere. The commutation relation obtained in Eq.~\eqref{Eq:Moduli-Commut} has the same form as the one defining our fuzzy sphere [see Eq.~\eqref{Eq:Space-Commut}], where the number of vortices on a sphere enters the expression in a similar way to the additional fuzzy scale, $N$. However, although this immediately points towards the fact that quantum vortices induce non-commutativity of space, the two commutators carry different physical meanings. Namely, vortices possess non-commutativity of their collective coordinates, but this does not directly translate to non-commutativity of spacial coordinates of the bosonic fields. Going one step further, we suggested a new definition of vortices existing on a fuzzy sphere. It accounts for a certain discreteness of space induced by non-commutativity, which  leads to the fact that point vortices become finite-width objects. This, however, should not be confused with an ordinary vortex core, since this shift from a point-like object to a finite-sized one is the direct result of the non-commutativity inherent to the fuzzy sphere construction.

As one of possible experimental realizations of a Bose system on a sphere, one could consider ultracold atoms on  a thin spherical shell trap, similar to one recently realized ~\cite{Shell-trap}. To induce fuzziness in this way, it would most likely be necessary to have a synthetic magnetic monopole field akin to the one in a Floquet engineering proposal~\cite{Synthetic-LLs_PhysRevLett.120.130402}, which aimed at realization of Haldane's construction for fractional quantum Hall states using a spherical geometry~\cite{Haldane_PhysRevLett.51.605,Fano_PhysRevB.34.2670,Greiter_PhysRevB.83.115129}. Alternatively, ultracold atoms can be used to engineer a bilayer superfluid state~\cite{Bilayer-superfluid}.  The latter is of interest since it offers an opportunity to study the behavior of a spherical system without actually realizing such geometry. This follows from the fact that a two-dimensional sphere can be defined by means of an atlas consisting of two stereographic projections to two different planes. Thus, some specific bilayer superfluid state can be equivalent to superfluidity on a fuzzy sphere. Aside from the ultracold atoms, another candidate to investigate this state is a quantum Hall bilayer structure, which is known to be equivalent to a neutral superfluid \cite{Wen-Zee_PhysRevLett.69.1811}. Experimentally such a system exhibits an interlayer Josephson-like tunneling \cite{Spielman_PhysRevLett.84.5808}, although the system is not a superconductor \cite{Wilczek_PhysRevLett.86.1833,Balents_PhysRevLett.86.1825,Stern_PhysRevLett.86.1829}. 
In this problem, non-commutativity is intrinsic due to the Landau level projection. The theory of fuzzy superfluidity presented here serves a framework that can be adapted to describe particular experimental setups.

\begin{acknowledgments}
	We would like to thank Jeroen van den Brink for interesting discussions after the results of this paper were presented in a workshop by one of us (V.S.). 
\end{acknowledgments}

\appendix
\section{Detailed calculation of the normal fluid density}
\label{App:rhon}

In order to calculate the normal fluid density $\rho_n$, it is useful to separate the sum in Eq. \eqref{Eq:rho-n} in two terms, 
\begin{eqnarray}
	\rho_n&=&\frac{T}{2}\sum_{n=-\infty}^{\infty}\sum_{l=0}^{M}(2l+1)\frac{\frac{l(l+1)}{R^2}}{\omega_n^2+\frac{v^2l(l+1)}{R^2}}
	\nonumber\\
	&-&T\sum_{n=-\infty}^{\infty}\sum_{l=0}^{M}(2l+1)\frac{\frac{l(l+1)\omega_n^2}{R^2}}{\left[\omega_n^2+\frac{v^2l(l+1)}{R^2}\right]^2}.
\end{eqnarray}
Introducing the variable, $x=v^2l(l+1)/R^2$, and treating $l$ approximately as a real variable rather than an integer, we rewrite the above as, 
\begin{eqnarray}
	\label{Eq:rhon-integrated}
	4\pi\rho_n&=&\frac{T}{2v^4}\sum_n\int_{0}^{\Lambda}dx\frac{x}{\omega_n^2+x}\nonumber\\
	&-&\frac{T}{v^4}\sum_n\int_{0}^{\Lambda}dx\frac{x\omega_n^2}{(\omega_n^2+x)^2},
\end{eqnarray}
where we have defined $\Lambda=v^2M(M+1)/R^2$. 
Performing the integration over $x$ for these two terms yields, 
\begin{equation}
	\int_0^\Lambda dx\frac{x}{\omega_n^2+x}=
	\Lambda-\omega_n^2\ln\left(\frac{\omega_n^2+\Lambda}{\omega_n^2}\right),
\end{equation}
\begin{equation}
	\int_0^\Lambda dx\frac{x\omega_n^2}{(\omega_n^2+x)^2}=
	\omega_n^2\ln\left(\frac{\omega_n^2+\Lambda}{\omega_n^2}\right)
	+\frac{\omega_n^4}{\omega_n^2+\Lambda}-\omega_n^2,
\end{equation}
so that the normal fluid density can be rewritten as, 
\begin{equation}
	\rho_n=
	\frac{1}{v^4}\left[
	\frac{\Lambda}{2}T\sum_n 1
	+\frac{3}{2}S_1(\Lambda)
	-S_2(\Lambda)+T\sum_n \omega_n^2\right],
\end{equation}
where for conciseness we used 
\begin{equation}
	S_1(x)=-T\sum_n \omega_n^2\ln\left(\frac{\omega_n^2+x}{\omega_n^2}\right),
\end{equation}
\begin{equation}
	S_2(x)=T\sum_n \frac{\omega_n^4}{\omega_n^2+x}.
\end{equation}
Differentiating,
\begin{equation}
	\frac{dS_1}{dx}
	=-T\sum_n \frac{\omega_n^2}{\omega_n^2+x}
	=xT\sum_n \frac{1}{\omega_n^2+x},
\end{equation}
and using, 
\begin{equation}
	T\sum_n \frac{1}{\omega_n^2+x}=
	\frac{1}{2\sqrt{x}}\coth\left(\frac{\sqrt{x}}{2T}\right),
\end{equation}
we obtain
\begin{equation}
	\frac{dS_1}{dx}=
	\frac{\sqrt{x}}{2}\coth\!\left(\frac{\sqrt{x}}{2T}\right).
\end{equation}
Thus, 
\begin{equation}
	S_1(x)=
	\int_0^{\sqrt{x}} dEE^2\coth\left(\frac{E}{2T}\right).
\end{equation}
Now we rewrite the above integral using, 
\begin{equation}
	\coth\left(\frac{E}{2T}\right)=1+\frac{2}{e^{E/T}-1},
\end{equation}
to obtain, 
\begin{equation}
	S_1(x)=\int_0^{\sqrt{x}} dEE^2
	+2\int_0^{\sqrt{x}} dE\frac{E^2}{e^{E/T}-1}.
\end{equation}
We have that, 
\begin{eqnarray}
	\label{Eq:Polylogs}
	\int dE\frac{E^2}{e^{E/T}-1}&=&T E^2\ln(1-e^{-E/T})
	-2T^2 E{\rm Li}_2(e^{-E/T})
	\nonumber\\
	&-&2T^3{\rm Li}_3(e^{-E/T}).
\end{eqnarray}
Differentiating the RHS above with respect to $E$ in Mathematica precisely yields the integrand in Eq. \eqref{Eq:Polylogs}.  

Therefore, the final result for $S_1(x)$ is, 
\begin{eqnarray}
	\!\!\!\! \! \!\!\! S_1(x)\!&=&\!\frac{x^{3/2}}{3}
	+2Tx\ln(1-e^{-\sqrt{x}/T})+4T^3\zeta(3)
	\nonumber\\
	\!&-&\!4T^2\sqrt{x}{\rm Li}_2(e^{-\sqrt{x}/T})
	-4T^3{\rm Li}_3(e^{-\sqrt{x}/T}).
\end{eqnarray}

For $S_2(x)$ we first write, 
\begin{equation}
	\frac{\omega_n^4}{\omega_n^2+x}=\omega_n^2-x+\frac{x^2}{\omega_n^2+x},
\end{equation}
which gives, 
\begin{equation}
	S_2(x)=T\sum_n \omega_n^2-xT\sum_n 1
	+x^2 T\sum_n \frac{1}{\omega_n^2+x},
\end{equation}
and hence, 
\begin{equation}
	S_2(x)=\frac{x^{3/2}}{2}\coth\left(\frac{\sqrt{x}}{2T}\right).
\end{equation}

Putting everything together leads to, 
\begin{equation}
	4\pi\rho_n=\frac{1}{v^4}
	\left[\frac{3}{2}S_1(\Lambda)-\frac{\Lambda^{3/2}}{2}\coth\left(\frac{\sqrt{\Lambda}}{2T}\right)
	\right],
\end{equation}
which is written more explicitly in Eq. \eqref{Eq:rhon-final}.

\bibliography{Fuzzy-SF-refs}

\end{document}